The Response of Atmospheric Chemistry on Earthlike Planets

around F, G and K Stars to Small Variations in Orbital Distance

John Lee Grenfell[1*], Barbara Stracke[1], Philip von Paris[1], Beate Patzer[2], Ruth Titz[1],

Antigona Segura[3,**], and Heike Rauer[1,2]

(1) Institut für Planetenforschung

Extrasolare Planeten und Atmosphären

Deutsches Zentrum für Luft- und Raumfahrt (DLR)

Rutherford Str. 2

12489 Berlin

Germany

(2) Zentrum für Astronomie und Astrophysik

Technische Universität Berlin (TUB)

Hardenbergstr. 36

10623 Berlin

Germany

(3) Department of Geosciences

443 Deike Building

Pennsylvania State University,

University Park, PA 16802, USA

* Corresponding author, lee.grenfell@dlr.de

**Presently at: Infrared Processing and Analysis Center

California Institute of Technology

770 South Wilson Avenue

Pasadena, CA 91125, USA

*Abstract:* One of the prime goals of future investigations of extrasolar planets is to search for life as we know it. The Earth's biosphere is adapted to current conditions. How would the atmospheric chemistry of the Earth respond if we moved it to different orbital distances or changed its host star? This question is central to astrobiology and aids our understanding of how the atmospheres of terrestrial planets develop. To help address this question, we have performed a sensitivity study using a coupled radiative-convective photochemical column model to calculate changes in atmospheric chemistry on a planet having Earth's atmospheric composition, which we subjected to small changes in orbital position, of the order of 5-10% for a solar-type G2V, F2V, and K2V star. We then applied a chemical source-sink analysis to the biomarkers in order to understand how chemical processes affect biomarker concentrations. We start with the composition of the present Earth, since this is the only example we know for which a spectrum of biomarker molecules has been measured. We then investigate the response of the biomarkers to changes in the input stellar flux. Computing the thermal profile for atmospheres rich in $H_2O$, $CO_2$ and $CH_4$ is however a major challenge for current radiative schemes, due, among other things, to lacking spectroscopic data. Therefore, as a first step, we employ a more moderate approach, by investigating small shifts in planet-star distance and assuming an earthlike biosphere. To calculate this shift we assumed a criteria for complex life based on the Earth, i.e. the earthlike planetary surface temperature varied between $0^oC < T_{surface} < 30^oC$, which led to a narrow HZ width of (0.94-1.08) Astronomical Units (AU) for the solar-type G2V star, (1.55-1.78) AU for the F2V star, and (0.50-0.58) AU for the K2V star. In our runs we maintained the *concentration of atmospheric $CO_2$ at its present-day level. In reality, the $CO_2$ cycle (not presently included in our model) would likely lead to atmospheric $CO_2$ stabilising at higher levels than considered in our runs near our quoted "outer" boundaries. The biomarkers $H_2O$, $CH_4$ and $CH_3Cl$ varied by factors 0.08, 17, and 16 respectively in the total column densities on moving outwards for the solar case. Whereas $H_2O$ decreased moving outwards due to cooling hence enhanced condensation in the troposphere, $CH_4$ and $CH_3Cl$ increased associated with a slowing in $H_2O+O^1D-->2OH$, hence less OH, an important sink for these two compounds. Ozone changes were smaller, around a 10% increase on moving outwards partly because cooler temperatures led to a slowing in the reaction between $O_3$ and $O^1D$. We also considered changes in species which impact ozone – the so-called family species (and their reservoirs), which can catalytically destroy ozone. Hydrochloric acid (HCl), for example, is a chlorine reservoir (storage) molecule, increased by a factor 64 in the mid-stratosphere (32km) on moving outwards for the solar case. For the F2V and K2V stars, similar sources and sinks dominated the chemical biomarker budget as for the solar case and column trends were comparable.*

*Keywords: Biomarkers, earthlike, exoplanets, atmosphere.*

## 1. Introduction

In this paper we discuss the variation of biomarker molecules such as $H_2O$, $O_3$, $CH_4$, $CH_3Cl$, and $N_2O$ on earthlike, extrasolar planets having an earthlike biosphere when the orbital distance is changed by a small amount and the central star class is varied (G2V, F2V, K2V). We start with the Earth as our most reliable reference and, using radiative-photochemical model, subject the atmosphere to small changes in order to learn about how terrestrial atmospheres could have developed.

Ultimately we wish to understand the development of terrestrial planet atmospheres over the entire Habitable Zone (HZ), commonly defined to be where liquid water exists on a planet's surface (*Rasool and DeBergh* 1970; *Hart*, 1978). However, achieving this with radiative-photochemical models is currently rather challenging. This is because near the inner-edge of the HZ water evaporates strongly leading to very damp atmospheres whereas near the outer-edge, high $CO_2$ or/and $CH_4$ atmospheres with strong greenhouse heating are possible. Calculating the thermal profiles in these cases is currently challenging. With this in mind, we have chosen to introduce only small changes of around 5-10% in the orbital distance of an earthlike biosphere orbiting different main sequence stars, providing reliable temperature profiles with our model. Although our experiments do not consider the full HZ range, they provide nevertheless valuable clues into how responses in atmospheric chemistry could influence the development of earthlike extrasolar planetary atmospheres.

Changes in atmospheric chemical processes are expected to influence significantly the abundances of biomarkers to be investigated by proposed satellite missions like Darwin or TPF. In this paper, we therefore focus on understanding the atmospheric chemistry of biomarker molecules. A comprehensive review of tropospheric and stratospheric chemistry is provided by the *World Meteorological Organization (WMO) Report*, (WMO, 1998; WMO, 2002). For the purposes of this paper we will review here the major chemical sources and sinks of important biomarker molecules. Note that the chemical processes described operate in the model and understanding them is directly relevant to explaining the biomarker response.

*Biomarker Photochemistry:* Ozone in the stratosphere is mainly derived from the photolysis of $O_2$ in the Herzberg continuum ($185<\lambda<242$nm). *Chapman* (1930) first formulated a chemical theory to account for the stratospheric ozone layer. Ozone is also destroyed by catalytic cycles involving $HO_x=(OH+HO_2)$ (*Bates and Nicolet*, 1950), $NO_x=(N+NO+NO_2)$ (*Crutzen*, 1970) and $ClO_x=(Cl+ClO)$ (*Molina and Rowland*, 1974). These represent the so-called active forms of hydrogen, nitrogen and chlorine respectively, because they can lead to ozone destruction. The stratospheric chemistry of ozone is summarised in Figure 1. The active forms may be de-activated via chemical reactions into passive forms, also called reservoir molecules. For example, nitric acid, $HONO_2$ (also written $HNO_3$) is a reservoir for the active forms, $OH$ and $NO_2$. The reservoir molecules are shown in bold near the outer edges of Figure 1 – they do not destroy ozone directly, but they may release the active form under other physical conditions, e.g. via photolysis or thermal decomposition. The reservoirs in Figure 1 have been studied fairly extensively. *WMO* (2002) provides a good overview of their composition profiles, sources, and sinks on the Earth.

Water is photochemically inert in the troposphere. Its entry into the stratosphere is controlled by the tropopause temperature, which acts as a cold trap for damp air rising from below. *Sherwood* (2001) has reviewed processes regulating the entry of $H_2O$ from the troposphere into the stratosphere. Another important source of stratospheric water is: $CH_4 + OH \rightarrow CH_3 + H_2O$. $H_2O$ is removed in the upper stratosphere and mesosphere via photolysis in the Lyman-alpha (121.6nm) band. Figure 2 summarises the photochemical sources and sinks of water. Minor sources are shown in a smaller typeface.

Methane is formed mainly from methanogenic bacteria under anaerobic conditions (wetlands and oceans) (*Intergovernmental Panel on Climate Change IPCC*, 2001; *Krüger et al.*, 2001) and via geologic activity (*Etiope and Klusman*, 2002). It photolyses only very slowly in the troposphere, where it is mainly removed by reaction with the hydroxyl radical (OH). OH is sometimes referred to as the detergent of the troposphere, because it removes via oxidative degradation a wide range of volatile organic carbon pollutants. In the stratosphere methane is removed via reaction with the OH radical and via photolysis in the upper stratosphere. OH is mainly formed via: $O_3 + h\nu \rightarrow O_2 + O^1D$ followed by $O^1D + H_2O \rightarrow 2OH$. Figure 3 summarises the atmospheric sources and sinks of methane.

$N_2O$ is emitted into the atmosphere mainly via denitrifying bacteria (e.g. *Oonk and Kroeze*, 1998). Bacterial emissions are sensitive to e.g. soil type, rainfall, and temperature (*Liu*, 1996). $N_2O$ is photochemically inert in the troposphere. In the stratosphere, about 95% of $N_2O$ is removed via: $N_2O + h\nu$ (<341nm) --> $N_2+O^1D$. The remaining 5% reacts with $O^1D$ to form NO. The sources and sinks of $N_2O$ are summarised in Figure 4.

Here we quantify the sensitivity of important biomarkers in atmospheres of exoplanets with earthlike biospheres around solar-type G2V, F2V, and K2V stars to modest radiation and temperature changes. We analyse biomarker changes in terms of shifts in their major chemical sources and sinks. For ozone, whose atmospheric chemistry is more complex than that of the other biomarkers, we also report chemical changes in the active species which destroy ozone, as well as changes in the associated reservoir molecules.  Our study assumes a modern day Earth composition. Clearly, this is a rather selective choice. However, it is the only example we have for which a full biomarker spectrum can be measured, e.g. by analysing Earthshine. The approach is well-established in the literature e.g. Woolf et al. (2002), Ford et al. (2005) measured modern-day Earthshine spectra "to pave the way for TPF", des Marais et al. (2002) assess (modern) Earth spectra to aid with TPF goals, Seager et al. (2005) analyse IR spectra of the modern Earth in connection with TPF habitability studies. Clearly, future studies have to be expanded e.g. by isolating the anthropogenic component and studying early earthlike atmospheres.

## 2. Computational details

### 2.1. Model description

     The original code has been described in detail by *Kasting et al.* (1985) and developed by *Segura et al.* (2003). The chemistry includes 55 species for 217 reactions from the surface up to 64 km at 1 km intervals. We updated the reaction rates from the *Jet Propulsion Laboratory (JPL) Report 14* (2003). The troposphere

10 includes methane oxidation as well as wet and dry deposition. The chemistry features $HO_x$, $NO_x$, $O_x$ and $ClO_x$ chemistry as well as their major reservoirs. Photolysis was diurnally averaged for clear skies. The chemical integrator employed the backward Euler method. As in *Segura et al.* (2005), the surface fluxes used for the biogenic compounds were intended to be those needed to produce their observed mixing ratios

15 in the present-day Earth atmosphere. This assumed that biological production of trace gases is the same as on Earth. This treatment is consistent with the conservative changes in orbital distance applied in this study.

     The radiation module employed 52 levels from the surface up to 70 km. The shortwave code was two-stream based on *Toon et al.* (1989) using the correlated-k

20 approach to parameterise absorption by the common greenhouse gases. The longwave code was based on the method of *Mlawer et al.* (1997), which uses a weighting method favouring wavelengths where absorption changes more rapidly.

The radiation and chemistry modules were run consecutively with coupling between the two modules via interpolation of temperature and radiative gas concentrations between their vertical domains such that output from one module was subsequently introduced as starting values into the other and the coupling process repeated until results converged.

Compared with *Segura et al.* (2003) we have updated the chemical reaction kinetics as mentioned before and implemented some code optimisation. The Segura work employed a wavelength-integrated surface albedo of 0.20 in order to attain the observed surface temperature of 288K. This is somewhat higher than the observed Earth value of 0.10; the difference is intended to compensate for missing cloud processes in the model. In our runs, the updates made required us to adopt a surface albedo value of 0.218 in order to achieve a surface temperature of 288K. Like in the Segura work, we then adjusted the surface methane emissions until we attained the observed tropospheric methane concentration of $1.6 \times 10^{-6}$ vmr. In this way we arrived at a new methane emission value of 582Tg/year, being closer to the observed Earth value of 535Tg/year (Houghton et al., 1995) than the previously adopted value in the model of 954 Tg/year.

**2.2 Parameter Variation**

All runs employed the same, initial, modern-day earthlike composition. For each main sequence star chosen, the stellar fluxes were then changed based on the determined position (see next paragraph) and assuming the appropriate spectrum. For the runs of the F2V star and K2V star we adopted the same flux-scaling procedure as discussed in *Segura et al.* (2003). There is a choice when defining the "centre" of the HZ around different stars for our control runs. One can take either (1) the distance where the incoming integrated stellar flux is the same as the Earth, i.e. placing the planet at the same relative position in the HZ as Earth for the solar case, or (2) the distance where the model calculates a mean surface temperature of 288K. For the Sun and the K2V cases the two definitions lead to similar star-planet distances (see below) but for the F2V case, definition (1) leads to 1.69AU whereas (2) (which we took in this work) leads to 1.65AU.

A series of runs were then performed in which the star-planet distance (hence the incoming stellar flux) was systematically decreased relative to the planetary position of a control run (at 1AU in the solar case) in units of 0.01AU. We then stopped when the surface temperature reached 30°C. This limit corresponds to an early "human habitability" definition of the HZ from *Dole* (1964), which was introduced before the full potential of extremophile life on Earth was recognised. Note that this orbital range is narrow compared with modern HZ definitions (e.g. Kasting et al., 1993), but, as we discussed it is challenging for our model to calculate the full range, so we concentrate instead on a reliable temperature range. Our 30°C surface temperature constraint was reached at an inner boundary of 0.94AU for the G2V star. A similar procedure to determine the outer orbital position boundary, where a surface temperature limit of 0°C was chosen, led to 1.08AU for the G2V case. Note that in reality, an "Earth" with a mean surface temperature of 0°C is unlikely, since the ice-albedo effect becomes important once the mean temperature falls below about 10°C, leading to rapid further cooling of the planet. Also, our model does not (yet) feature a $CO_2$ cycle response and we maintain atmospheric $CO_2$ at a constant level. So our quoted outer boundary value (e.g. 1.08AU for the the G2V case) is an over-estimate for a planet with a constant level of $CO_2$ and no other greenhouse gases.

In this way, we identified the following orbital positions, over which we discuss the development of atmospheric chemistry of biomarker molecules for the different main sequence stars. Values in brackets show the % distance from the control run:

Solar-type G2V star:

Run (1): Inner edge 0.94 AU (6.0%), $T_{surf}$=302.9K

Run (2): Solar-Control, 1.00 AU, $T_{surf}$=288.1K

Run (3): Outer edge 1.08 AU (8.0%), $T_{surf}$=272.8

F2V star:

Run (4): Inner edge, 1.55 AU (6.1%), $T_{surf}$=302.5K

Run (5): F2V-Control, 1.65 AU, $T_{surf}$=288.3K

Run (6): Outer edge 1.78 AU (7.9%), $T_{surf}$=273.3K

K2V star:

Run (7): Inner edge, 0.50 AU (5.7%), $T_{surf}$=302.2K

Run (8): K2V-Control, 0.53 AU, $T_{surf}$=289.3K

Run (9): Outer edge, 0.58 AU (9.4%), $T_{surf}$=272.1K

## 3. Results and Discussion

### 3.1 Temperature Profiles

Figure 5 shows temperature profiles for the inner boundary run (solid), mid boundary (dashed) and outer boundary run (dotted) for (a) solar G2V star, (b) F2V star, and (c) K2V star. For all three cases, there featured strong cooling throughout the troposphere, on moving away from the central star. Also in all three cases, the height of the tropopause, where temperature started to increase with height, was consecutively lowered, suggesting that tropospheric air contracted in response to the cooling. A similar effect was calculated for the height of the stratopause. In the stratosphere, for the G2V and F2V stars, there did not feature a strong temperature response with varying orbital distance. In this region of the atmosphere, radiative heating mainly from ozone and $CO_2$ dominate the heat budget. $CO_2$ was invariable in our model runs and the stratospheric ozone profile did not change greatly with varying orbital distance. The weak ozone response was related to opposing effects; near the outer orbital region, lower temperatures implied an ozone *increase* due to temperature-dependent ozone (Chapman) photochemistry, whereas the increase in $HO_x$, $ClO_x$ and $NO_x$ (discussed below) implied an ozone *decrease*. We discuss shortly the chemistry effects in more detail.

An interesting effect occurs above 50km for both the F2V (Figure 5b) and K2V (Figure 5c) cases, whereby the inner run (solid line) is actually colder than the mid run (dashed line). We suggest that this was related to the differing responses of the radiatively active gases to the different UV environments of the various stars. The effect is treated in more detail below, where the chemical profiles of biomarkers are discussed.

**3.2 Total column (tropospheric + stratospheric) of biomarkers**

Figure 6 shows column densities for the biomarker molecules $O_3$, $CH_4$, $H_2O$, $N_2O$, and $CH_3Cl$ with varying orbital distance around typical main sequence stars: (a) the Sun (G2V), (b) F2V, and (c) K2V. Figures 6a-c suggest that the column trends remained the same for the various stars with varying orbital distance. The source/sink analysis also implied that similar processes dominated the biomarker chemical budget as for the solar case, so the F- and K-star results are discussed only briefly. Although the trends showed many similarities, absolute amounts were not the same. For example, the K star exoplanet, with its weak UV-B (290-320nm) environment, featured only a modest ozone layer which varied from 209 DU to 236 DU with increasing orbital distance. At the same time, the weaker UV-B environment however contributed to rather low OH concentrations, which resulted in a large methane column 3-6 times thicker than in the solar case. We now consider biomarker changes individually.

**Ozone** – column values in Figure 6 increased modestly, e.g. for the solar case from 296 DU for the inner run to 327 DU for the outer run. The increase was favoured e.g. by a lowering in temperature on moving outwards due to a slowing in the reaction $O+O_3 \rightarrow 2O_2$. However, catalytic cycles can also affect ozone, which we now consider.

**Families affecting ozone** - the so-called "family" species $HO_x$ (=$OH+HO_2$), $NO_x$ (=$NO+NO_2$), and $ClO_x$ (=$Cl+ClO$), can destroy ozone via chemical catalytic cycles. It is convenient to consider the sum of the species shown in brackets, because they rapidly interconvert with each other and it is the sum which is the important quantity affecting ozone. Model calculations in the mid-stratosphere near 30km, where the ozone concentration peaks, implied $NO_x$ increased by about 29%, $HO_x$ increased by 73% and $ClO_x$ increased by up to a factor of 8.1 on moving orbital distance outwards for the G2V case. Clearly, these changes suggest a decrease in ozone. Opposing this, the lowering in temperature as we move outwards suggests the opposite effect, as already mentioned. Such opposing effects may explain why the ozone concentration changed by only a small amount compared with the other biomarkers. The $NO_x$ increase on moving outwards was strongly favoured by an *increased* photolysis rate of $N_2O$ ($P_{N2O}$), an important source of $NO_x$, in the mid to upper stratosphere, where

$$P_{N2O} = j[N_2O] \text{ molecules cm}^{-3} \text{ s}^{-1}$$

where j=photolysis coefficient ($s^{-1}$) and square brackets denote concentration in molecules $cm^{-3}$. Near e.g. 30km $P_{N2O}$ increased from 538 molecules $cm^{-3}$ $s^{-1}$ for the inner run to 716 molecules $cm^{-3}$ $s^{-1}$ for the outer solar run. Further investigation revealed that the radiative flux in the lower to mid-stratosphere actually *increased* with increasing star-planet distance. On the upper levels we obtained the expected decrease in flux. However, on the lower levels the flux increased due e.g. to the strong lowering in the $H_2O$ column (discussed later). The increase in $P_{N2O}$ was also related to a large increase in $[N_2O]$ with increasing star-planet distance (see $N_2O$ section below).

The $HO_x$ increase for the runs with increasing orbital distance was related to changes in the reaction $H_2O+O^1D \rightarrow H_2O$, an important source of $HO_x$. Although $H_2O$ *decreased* strongly in the column on moving outwards, due to increased condensation in the troposphere (discussed below), in the stratosphere $H_2O$ *increased*. This was associated with increased stratospheric methane (see discussion below and Table 1), which oxidises in the stratosphere to form $H_2O$. Near e.g. 30km $H_2O$ increased from $4.8 \times 10^{-6}$ to $9.0 \times 10^{-6}$ by volume mixing ratio for the solar control and outer runs respectively.

The $ClO_x$ increase with increasing orbital distance was favoured by an increase in the reaction $HCl+OH \rightarrow Cl+H_2O$ – this was an important source of $ClO_x$, which increased with orbital distance mainly because the HCl reservoir, which e.g. near 30km for the solar case increased by a factor 64 (discussed below).

**Reservoirs affecting ozone** - No discussion of modelling terrestrial ozone columns is complete without consideration of so-called chemical "reservoir" molecules. These are so-named because they act as storage agents or "reservoirs" for the active family forms which can destroy ozone, as discussed in the previous section. In the models, it is useful to know the starting values of reservoir compounds, in order to make a good estimation of ozone concentrations. Although global observation datasets via satellite measurements of reservoir molecules are only just beginning to be compiled for the Earth, recent progress has been rapid, and, should these species be present in high abundances on exoplanets, it is feasible that measurements for them may be attempted in the not-too-distant future. We have investigated the response of a variety of reservoir molecules ($HCl$, $H_2O_2$, $HNO_3$, $ClONO_2$). Of these, $HCl$ and $ClONO_2$ gave the largest responses, so they are discussed here further.

HCl values were about 64 times higher for the outer run compared with the inner run and about 6 times higher for the outer run than for the solar control run. HCl is formed in the stratosphere mainly via $Cl+CH_4 \rightarrow CH_3+HCl$. In the upper stratosphere, nearly all inorganic chlorine exists as HCl. It is destroyed via transport to the troposphere followed by rainout, by reaction on the surface of sulphate aerosol or polar stratospheric clouds, or via direct reaction with OH. *Beaver and Russell III* [1998] have discussed the global climatology of HCl. In the stratosphere, i.e. above about 20km, the increase in HCl for the outer runs was related to an enhancement in the reaction via $HO_2+Cl \rightarrow HCl+O_2$ whose rate increased at colder temperatures with a dependence $\exp(170/T)$, and also more methane (discussed below) hence faster $Cl+CH_4 \rightarrow HCl+CH_3$.

$ClONO_2$ increased by a factor 1.74 (near 30km where its concentration peaks) in the solar control run compared with the inner run and increased by a further factor 2.5 for the outer run in comparison with the corresponding control results. $ClONO_2$ is formed via $ClO+NO_2+M \rightarrow ClONO_2+M$ and is sensitive to photolysis and by reactions on the surface of sulphate aerosol and polar stratospheric clouds. Some stratospheric observations of $ClONO_2$ have been made for the Earth, e.g. *Sen et al.* [1999] although a global climatology is not available. The increase in $ClONO_2$ with increasing orbital distance was associated partly with the $ClO_x$ increase as already discussed and also partly due to a slower photolytic sink.

**Methane** - the methane column in Figure 6 increased strongly with increasing orbital distance for all three main-sequence stars considered, e.g. varying from 327 DU (inner), 1234 DU (control) up to 5679 DU (outer), resulting in an overall increase by a factor of 17. To understand this change, we inspected the rates of all methane sinks in the troposphere at 10km, i.e. where the methane column contribution was significant. Note, that there features no in-situ chemical sources of methane in the model, only a biogenic surface mass flux. For e.g. in the solar control run the fastest methane sink for all runs was via reaction with OH, which at 10km proceeded about 86 times faster than the next-fastest sink, namely reaction with Cl ($Rate_{CH4+OH}=1.2 \times 10^4$ molecules $cm^{-3}$ $s^{-1}$, $Rate_{CH4+Cl}=1.4 \times 10^2$ molecules $cm^{-3}$ $s^{-1}$). The rate of the reaction of methane with OH, $Rate_{CH4+OH}=k[OH][CH_4]$ dropped from $4.5 \times 10^4$ molecules $cm^{-3}$ $s^{-1}$ (inner run), to $1.2 \times 10^4$ molecules $cm^{-3}$ $s^{-1}$ (outer run ), i.e. a total decrease by a factor of 3.4. We now consider each term in turn in order to understand more fully the methane response. The rate constant, k varies as $\exp(-1775/T)$, hence decreased by a factor 5.8 at 10km on moving outwards, since temperature decreased from 257K (inner run) to 205K (outer run).

For the solar case OH decreased by a factor 10 in vmr from the inner to the outer runs at 10km. OH is formed mainly via $jO_3 \rightarrow O^1D$ then $O^1D + H_2O \rightarrow 2OH$. On moving outwards the troposphere becomes much drier due to reduced evaporation which suggested a decrease in OH.

In summary: $R_{CH4+OH}$ (decreases by factor 3.4 over HZ) = k (decreases by factor 5.8 on moving outwards) * [OH] (decreases by a factor 10) * [$CH_4$]
Balancing sides implies that methane must have increased by a factor, $f=(5.8*10/3.4) = 17$, at 10km on moving outwards, which indeed happened in the model.

**Water** - water columns in Figure 6 decreased strongly with increasing orbital distance e.g. from $8.4 \times 10^6$ DU (inner run, solar case), to $2.8 \times 10^6$ (control) and to $0.7 \times 10^6$ DU (outer run). The convection routine in the model assumed a moist adiabat (*Ingersoll*, 1969; *Ingersoll et al.*, 2000). The water column decreased with increasing orbital distance, because its value is dominated (>99%) by the tropospheric component. Here, $H_2O$ photochemistry was negligible, and the overriding effect was a strong sensitivity of moisture to ambient temperature via condensation. The magnitude of the temperature decrease (e.g. by 52K at 10km) and the corresponding factor of twelve decrease in the column was consistent with typical established temperature-% Relative Humidity (RH) Relations in the troposphere, e.g. *Strahler*, (1963).

**Nitrous oxide** - the nitrous oxide column in Figure 6 increased with increasing orbital distance from 209 DU (inner run), 231 DU (control) up to 258 DU (outer run). As for methane, we output $N_2O$ sources and sinks in the mid-troposphere at 10km, since about 95% of the $N_2O$ column occurs in the troposphere. In the model, $N_2O$ featured a biogenic source from the surface and a weak, inorganic chemical source: $N_2+O^1D \rightarrow N_2O$.

Could the inorganic source lead to a false positive detection of $N_2O$ as a biomarker? At 10km this source proceeded very slowly, between R=0.4-0.6 molecules cm$^{-3}$ s$^{-1}$ in the solar runs. This seems insufficient for $N_2O$ to build-up to significant amounts in about 150 years, which is a typical $N_2O$ lifetime against its atmospheric sinks. Over this period, an atmosphere featuring such an inorganic source only would (ignoring sinks) produce about $2.4 \times 10^9$ $N_2O$ molecules cm$^{-3}$ at the surface, which is about 3000 times lower than if we include the biogenic fluxes.

$N_2O$ chemistry could not explain the afore-mentioned increase in $N_2O$ on moving outwards across our HZ. The explanation was related to changes in mixing. Cooling and an increase in density in the outer run column led to a lowering of the Eddy diffusion coefficient, which displays a weak dependence upon density. This led to $N_2O$ preferentially remaining on lower levels for the outer run.

**Methyl chloride** ($CH_3Cl$) – the column value in Figure 6 increased from 0.1 DU (inner run), to 0.4 DU (control) and to 1.6 DU (outer run) with increasing orbital distance of the G2V star. $CH_3Cl$, like $CH_4$ has only a biogenic source in the model. It

has three sinks, namely:

$$CH_3Cl + OH \rightarrow Cl + H_2O$$

$$CH_3Cl + Cl \rightarrow Cl + HCl$$

$$CH_3Cl + h\nu \rightarrow CH_3 + Cl$$

At 10km our analysis implied that the OH sink was faster than the Cl sink by a factor f, where f=83 (inner solar run), f=318 (outer solar run), so we consider only the reaction with OH in our analysis here. The rate constant of the OH sink reaction had a positive temperature-dependence, which varied as exp(-1250/T). The reduced temperature for the outer run led to a slowing in this important sink, which accounted in part for the biomarker decrease.

**3.3 Stratospheric column of biomarkers across the solar HZ**

Planets with thick (>5bar) atmospheres and which rotate slowly (e.g. Venus) are expected to feature large-scale Hadley cell circulations with dayside heating in the tropics leading to strong upwelling, which favours cloud formation, with concomitant sinking at higher latitudes and return of air equatorwards on lower levels. Such planets may well feature an ubiquitous, thick mantle of clouds up to the tropopause, in which case observations will likely be restricted to the stratosphere and above. With this in mind, it is interesting to extract the stratospheric components of our modelled column values and to investigate how these vary with changing orbital distance. The stratosphere was defined to begin where the temperature profile began to increase with altitude, which in the model for the solar runs corresponds to 23km for the inner run, 18km for the control, and 16km for the outer run (cf. Figure 5a). These values demonstrate that the tropopause height decreased with increasing orbital distance, which arose due to cooling and contraction of air. Table 1 illustrates the stratospheric column contribution of the biomarkers of the G2V star.

Table 1 suggests that the stratospheric ozone column changed relatively little with changing orbital distance compared with the other biomarkers. As already discussed, the decrease in UV and increase in ozone-destroying families implying that a decrease in ozone with increasing orbital distance was opposed by the temperature decrease, which implied the opposite effect by the Chapman mechanism.

Methane values in Table 1 increased from 9 DU to 301 DU with increasing orbital distance. About 90% of this increase was attributable to the lowering in altitude of the tropopause, leading to a thicker stratospheric column. Changes in the chemistry accounted for the remaining 10% change. A chemical analysis at 50km for the solar run suggested that OH, unlike in the troposphere, was no longer the main sink for methane. Reaction with $O^1D$ accounted for 49% of the chemical loss, OH 33% and Cl about 18%. Colder temperatures with increasing orbital distance led to a lowering in the rate constants, $k_{(CH4+OH)}$ by a factor 2.5, and $k_{(CH4+Cl)}$ by a factor 2.0 (at 50km) which favoured the stratospheric methane increase.

The stratospheric water column in Table 1 is very low compared with its tropospheric counterpart. As in the troposphere, it decreases by a factor of about 3 from the inner to the outer solar run, which we interpret as the drier troposphere resulting in less transport of water across the cold trap.

Stratospheric $N_2O$ values in Table 1 increased by a factor of about two for the outer compared with the inner runs. The outer run featured a denser and cooler column which led to a small lowering in Eddy diffusion. This favoured a downward shift in $N_2O$, whose only source of importance in the model is biogenic emission from the ground. This in turn led to overall weaker photolytic destruction for the outer run, hence the increase in $N_2O$ concentration.

$CH_3Cl$ values in Table 1 increased in the stratospheric column with increasing orbital distance partly due e.g. to a lowering in the photolytic sink in the mid to upper stratosphere.

**Biomarker ratios with changing orbital distance from main sequence stars**

It is interesting to inspect ratios of biomarkers because these quantities may respond more sensitively to changing orbital distance than the biomarkers themselves. Studying biomarker ratios may therefore provide an additional means of evaluating and comparing the models. The ratios also provide an indication of whether the biomarkers reside mainly in the stratosphere or troposphere so may provide clues to atmospheric structure or/and clouds. Table 2 accordingly shows the variation in the modelled ratios of column ($O_3/H_2O$) and ($CH_4/H_2O$) for the G2V, F2V, and K2V cases:

In Table 2 the ratio of total column ($O_3/H_2O$) was generally much lower than the stratosphere-only ratio. This arose because whereas most ozone resides in the stratosphere, most water resides in the troposphere.

On increasing orbital distance the total column ratio ($O_3/H_2O$) in Table 2 increases for the solar case by a factor 10. This was mainly due to colder temperatures drying out the troposphere with a smaller effect from weaker ozone photolysis leading to somewhat more ozone on moving outwards. The ratio of stratospheric column ($O_3/H_2O$) increased by a factor 4.5 for the solar case on moving outwards. This was less than the total column case mainly due to stratospheric water actually increasing with increasing orbital distance , because methane, its stratospheric precursor increased, as already discussed (see also Figure 7).

The total column ($CH_4/H_2O$) in Table 2 was much lower than the stratospheric ratio. Both $CH_4$ and $H_2O$ have the bulk of their column densities in the troposphere, but this is more extreme in the case of water, which is very effectively removed above the tropopause by the cold trap. So, similar to ($O_3/H_2O$), comparing the modelled ratio of ($CH_4/H_2O$) in Table 2 with future observations, should give an indication of whether a cloud-base is present at the troposphere. The total column ratio of ($CH_4/H_2O$) increased by a factor of 200 for the solar case on increasing orbital distance. This was due to an increase in $CH_4$ (due to less OH associated with weaker insolation and a drier atmosphere) and less $H_2O$ due to enhanced condensation. The stratospheric column ratio, ($CH_4/H_2O$) increased by a factor 110 on moving outwards (refer to absolute changes in stratospheric methane and water, discussed above). It should be noted, however, that the usefulness of methane as a biomarker, is somewhat constrained by the fact that there exist volcanic sources on the Earth, although they are weak.

**3.4 Profiles of biomarkers for the solar case**

Figure 7a-d show profiles of the biomarkers ozone, water, methane, and nitrous oxide respectively for the solar case. Ozone changes are relatively small (Figure 7a) so are not discussed further here. The water profile (Figure 7b) shows a large tropospheric increase for the inner run (plain line) due to enhanced evaporation, but in the stratosphere, it is the outer run which featured most water (dotted line in Figure 7b). To understand this, one must realise that methane oxidation is the main source of water in the stratosphere, and note that methane increased strongly throughout the column (Figure 7c). The bulk of the methane column increase occurred in the troposphere, where it was related to a lowering in its main sink, via reaction with OH, in turn due to lowered photolysis and also due to drying in the outer run as already discussed. Nitrous oxide (Figure 7d) increased in the troposphere with increasing orbital distance. This was mainly due to changes in mixing over the column related to shifts in density, as already discussed.

Section 3.1 drew attention to the fact that, above 50km, the inner runs were actually *colder* than the mid runs for the F2V and K2V cases. To investigate this further, we prepared plots analogous to Figure 7a-d but for the F2V and K2V cases (not shown). For the F2V case, the strongest change in the relevant region came from water. For the inner F2V run, the strong F2V UV-B environment led to almost complete oxidative loss of methane into water on the upper levels. This run was in particular extremely damp above 50km, where water reached (30-35) ppmv i.e. about 6 times damper than the solar case. This led to strong stratospheric greenhouse cooling, which probably played a large role in the anomalously cool inner run, already noted in the temperature profile. The temperature behaviour reflects opposing processes – moving inwards across the HZ, higher photon density leads directly to a *heating* effect. However, faster methane oxidation also leads to more water in the stratosphere, which is a strong greenhouse *cooler* at these levels. For the solar case, the first process appears to dominate the temperature profile, whereas for the F2V case, the stronger methane oxidation effect appears to lead to the second mechanism dominating.

For the K2V case (not shown) the UV-B environment was now *weaker* than that of the Earth but nevertheless featured, as for the F2V case, a cooler inner run than the mid run above about 50km. An important response in the K2V case came from ozone. The weak UV-B environment led to slow methane oxidation into water, hence low HOx. With little HOx to destroy it, K2V ozone climbed to about 6 times its concentration compared with the solar case above 50km. However, the methane and ozone responses, although quite large, could not explain the temperature behaviour since (1) the mid run featured more methane (and $CH_3Cl$) than the inner run i.e. suggesting a cooler mid run - not the response seen, and (2) the mid run featured less ozone than the inner run i.e. again suggesting a cooler mid run. The K2V temperature response could be partly explained by the water response, i.e. analogous to the mechanism explained above for the F2V case, i.e. the inner K2V run featured faster methane oxidation hence more water, which cooled the stratosphere more than the mid K2V run. However, the effect was less clear-cut than for the F2V case.

**4. Concluding remarks**

We took an earthlike biosphere and performed a sensitivity study varying orbital distance by (+/-) 5-10% in order to investigate the response of atmospheric chemistry. We did not consider the full range of the HZ because it is currently challenging to accurately model temperature profiles of the water-rich (inner HZ) and $CO_2$ rich (outer HZ) atmosphere although this is a task which we are currently working on.

We then looked at changes in biomarkers. Based on the assumption that biomass activity did not change, input mass fluxes of biogenic gases into the atmosphere were held constant. Despite this, our sensitivity study has showed that biomarkers varied by sometimes surprisingly significant amounts across the HZ.

The coupling between photochemistry and climate can lead to some interesting effects. e.g. in the lower stratosphere the radiative fluxes actually *decreased* on increasing star-planet distance for the solar case. This was partly due to the strong decrease in the water column for the outer run.

With increasing orbital distance total overhead column ozone *increased*
modestly, partly associated with a cooling effect. Total column water *decreased* by a
factor 12 (for the solar run) due to enhanced condensation in the troposphere. This led
to a slowing in the reaction $H_2O+h\nu \rightarrow 2OH$, hence a reduction in OH in the
troposphere. This in turn led to methane and $CH_3Cl$ *increasing* by factors 17 and 16
respectively in the total column for the solar case.

Ratios of biomarker concentrations appear to be very sensitive to the orbital
position and may provide hints as to biomarker profiles and the presence of clouds
(see next point) . The ratio of total column $(O_3/H_2O)$ and $(CH_4/H_2O)$, for example,
increased by factors of 10 and 200 respectively with increasing orbital distance for the
solar case (Table 2).

If an exoplanet were to feature a thick cloud-base near the tropopause, then
only the stratospheric column contribution would likely be measured. Typical values
for the ratios $(O_3/H_2O)$ and $(CH_4/H_2O)$ are very different depending on whether we
include the total column, or only the stratospheric component (see Table 2).
Therefore, the ratio may vary considerably depending on whether a cloud-base is
present near the tropopause.

There are caveats concerning the validity and applicability of our experiment.
Our approach assumes that, on moving across the HZ, biogenic output remains
terrestrial and constant, i.e. methane, chloromethyl and nitrous oxide emissions do not
change. In reality, the fauna and flora would very likely adapt to the differing
conditions, there would be a shift in speciation, hence some changes in biogenic,
radiative gas emissions, which may influence climate hence the width and position of the HZ. However, although individual living species may be different, we expect that a biosphere with broadly similar overall fluxes is possible. Useful in addressing these questions would be to couple a vegetation model to our current photochemical-climate column model. This is beyond the scope of this work, but is desirable for future studies.

Although the biomarker changes calculated in this work are significant in terms of changes normally observed in Earth Science, it is nevertheless not clear whether the low-resolution spectra proposed for missions such as Darwin / TPF would be able to record the changes suggested in our model. In future work, we will develop our model to be able to consider the full range of terrestrial atmospheres possible over the modernly-defined HZ. We will also calculate low-resolution spectra based on our model output.

**Acknowledgements**

We are grateful to Dr Alain Léger and two other anonymous referees for useful comments. We are also grateful to Dr James Kasting for providing original code and for useful discussion. This study was supported by the International Space Science Institute (ISSI) and benefited from the ISSI Team ``Evolution of Habitable Planets''. A. S. thanks the NASA Astrobiology Institutes Virtual Planetary Laboratory Lead Team, supported by the National Aeronautics and Space Administration through the NASA Astrobiology Institute under Cooperative Agreement Number CAN-00-OSS-01.

**Figure Captions**

Figure 1: Stratospheric chemistry of ozone. Shown is a schematic overview of the major stratospheric chemical sources and sinks. Reservoir molecules are shown in bold. "hv" denotes photolysis, "M" is a third-body required to carry away vibrational energy of the reaction, the triangle symbol denotes thermal decomposition. For HCl in the lower right, "rainout" denotes a sink involving transport from the stratosphere to the troposphere followed by removal by wet deposition.

Figure 2: Stratospheric chemistry of water.

Figure 3: Stratospheric chemistry of methane.

Figure 4: Stratospheric chemistry of nitrous oxide.

Figure 5: Temperature profiles for (a) the Sun (G2V), (b) an F2V star, and (c) a K2V star. The modelled profiles are shown for the inner orbital position (solid), mid orbital position (control) (dashed), and outer orbital position (dotted) for all stellar types considered.

Figure 6: Biomarker column values (Dobson Units, DU) for the inner, mid and outer runs for (a) the Sun, (b) an F2V star, and (c) a K2V star for the five biomarkers $O_3$, $CH_4$, $H_2O$, $N_2O$ and $CH_3Cl$. Note, that the K2V values are plotted on a different scale.

Figure 7: Biomarker concentration profiles shown in ($1\times10^6$) volume mixing ratio (vmr) for the solar case. Shown is the inner run(solid), mid run (dashed), and outer run(dotted) for (a) ozone, (b) water, (c) methane, and (d) nitrous oxide.

Table 1: Stratospheric column amount (DU) of various biomarkers. The values represent the available, measurable quantities for an earthlike exoplanet with an opaque cloud-base near the tropopause.

| Run | $O_3$ | $CH_4$ | $H_2O$ | $N_2O$ | $CH_3Cl$ |
|---|---|---|---|---|---|
| Inner | 201 | 9 | 931 | 5 | $1\times10^{-3}$ |
| Control | 215 | 52 | 340 | 8 | $1\times10^{-2}$ |
| Outer | 241 | 301 | 270 | 12 | $6\times10^{-2}$ |

Table 2: Variation in the biomarker ratios ($O_3/H_2O$) and ($CH_4/H_2O$) with varying orbital distance. Values are shown for the total column and for the stratospheric column component for each stellar type.

| Ratio | Sun | | F2V | | K2V | |
|---|---|---|---|---|---|---|
| | Inner | Outer | Inner | Outer | Inner | Outer |
| ($O_3/H_2O$) Total | $4 \times 10^{-5}$ | $4 \times 10^{-4}$ | $6 \times 10^{-5}$ | $7 \times 10^{-4}$ | $3 \times 10^{-5}$ | $3 \times 10^{-4}$ |
| ($O_3/H_2O$) Strat. | 0.2 | 0.9 | 0.1 | 1.3 | 0.4 | 0.7 |
| ($CH_4/H_2O$) Total | $4 \times 10^{-5}$ | $8 \times 10^{-3}$ | $3 \times 10^{-5}$ | $6 \times 10^{-3}$ | $1 \times 10^{-4}$ | $5 \times 10^{-2}$ |
| ($CH_4/H_2O$) Strat. | $1 \times 10^{-2}$ | 1.1 | $2 \times 10^{-3}$ | 0.8 | 0.1 | 6.4 |